# Resistance of high-temperature superconducting tapes triggered by alternating magnetic field

Quoc Hung Pham, Rainer Nast and Mathias Noe

*Abstract*—Dynamic resistance occurs in a superconducting tape carrying a dc transport current while being exposed to an alternating magnetic field. This effect is caused by flux movements interacting with the transport current. The dynamic resistance is already applied in many superconducting applications, for example superconducting flux pumps or persistent current switches. The resistance is highly dependent on the magnetic field and the frequency the superconductor is subjected to and its properties. When the dynamic resistance exceeds a certain value and thus enters the magnitude of the resistances of the normal conducting layers of the HTS tape, these normal conducting layers play a significant role in the total resistance of the tape. In this paper, modifications were made to the silver stabilizer and the total resistance of the HTS tape has been investigated. The experimental results with frequencies up to 1000 Hz and magnetic field up to 277 mT show significant increases in resistance. Additionally, a multilayer model based on H-formulation is presented to calculate the losses of the superconductor. The results also show significant heating due to the losses and therefore a temperature rise, which effects the measured total resistance. These results can be further used for applications where high switchable resistances are required with zero dc resistance when the magnet is turned off.

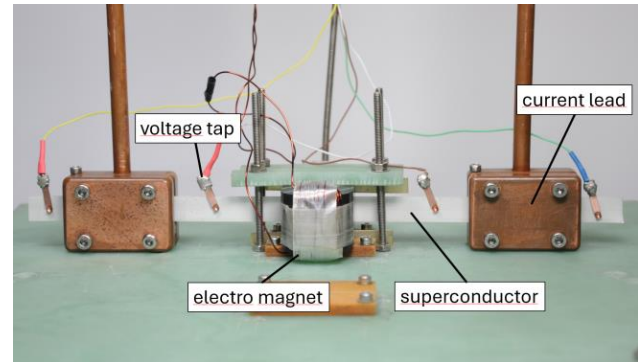

**Fig. 1.** Picture of the experimental setup

## I. INTRODUCTION

The transition from normal to a superconducting state can be triggered by current, temperature, or magnetic field. Exceeding the critical current of the superconductor is used for fault current limiters in electric grids while having nearly no impact on the impedance of the system in normal operation [1], [2], [3], [4], [5], [6].

The same transition can be achieved by heating the superconductor above the critical temperature. This is used for disconnecting superconducting coils from their power supplies leading to a persistent mode operation. Since heating is involved, thermal masses and heat transfer properties must be considered leading to switching times of several seconds [7], [8], [9], [10], [11].

For applications, where fast switching of superconductors is needed, the effect of dynamic resistance can be utilized by applying an alternating magnetic field to the superconductor. This has been applied in flux pumps [12], [13], [14]. Several papers on modelling and measuring the dynamic resistance with HTS tapes have been published [12], [14], [15], [16], [17], [18], [19], [20], [21], [22], [23], [24], [25], [26], [27], [28], [29], [30], [31], [32]. These previous works only investigated the dynamic resistance at lower frequencies and/or lower external magnetic field amplitudes, typically up 100 mT and 300 Hz.

We focus in this paper on a high overall resistance of the tape and therefore we investigate various configurations at higher frequencies and higher magnetic field amplitudes. The main idea in this paper is to use the total resistance of the tape to switch currents between parallel paths. The time needed for full current redistribution to another parallel conductor is mainly dependent on the built-up resistance of the switch in off-state. The higher the resistance is, the faster the commutation time. Therefore, in this work, the dynamic resistance is investigated over a wide range of frequencies and magnetic field amplitudes for several HTS tape configurations with modifications to the normal conducting layers, to increase the total resistance as far as possible.

## II. EXPERIMENTAL SETUP

### A. Description

The experimental setup is a basic apparatus for measuring critical currents of superconducting tapes with the addition of an iron core electromagnet to apply an alternating magnetic field. The maximum field is 300 mT and the maximum frequency is 1000 Hz. There are two sets of voltage taps contacted to the tape measuring the voltage with and without the current leads. The distance of the inner voltage taps is d = 10 cm and for the outer d = 26 cm. The voltages are recorded differentially by the data acquisition system NI USB-6281. The

Q. H. Pham is with the Karlsruhe Institute of Technology, Karlsruhe, Germany (e-mail: quoc.pham@kit.edu).
M. Noe is with the Karlsruhe Institute of Technology, Karlsruhe, Germany (e-mail: mathias.noe@kit.edu).
R. Nast is with the Karlsruhe Institute of Technology, Karlsruhe, Germany (e-mail: rainer.nast@kit.edu).

measurement range is set to ±0.1 V and the sampling rate to 10 kHz. The DC transport current via the current leads is supplied by a Keysight RP7943A power supply. The electromagnets are powered by the bipolar power supply Kepco BOP 72-6 which is controlled by an arbitrary waveform generator HP 33120A. The electromagnet consists of two horseshoe-shaped laminated iron cores, each of which is wound with enamel-magnet wire with a diameter of 0.75 mm. Each iron core has 50 windings. The whole experimental setup is submerged in liquid nitrogen at pool-boiling conditions. The dynamic resistance is then measured for a constant DC transport current of 3 A while varying the applied magnetic field from 0 mT to 277 mT and 500 Hz to 1000 Hz.

*B. Superconductor*

The investigated superconductor has the model number SF12100 and is manufactured by SuperPower. With a width of 12 mm, it has a critical current of 380 A at 77 K, self-field. The parameters of the tape are shown in Table 1. This tape has a silver layer of 1.5 µm on both sides and no copper stabilizer. To determine the total resistance of the tape, an equivalent circuit with resistors of the multilayer superconductor as shown in Figure 2 is used. The influence of the buffer layers is neglected due to their high resistance. Layer-to-layer resistances are also neglected due to the large current feed-in area of 12 mm x 50 mm.

According to the equivalent circuit, an applied transport current $I_t$ is divided between the different layers. The current flows according to the ratio between the resistances. The total resistance $R_{tot}$ can be calculated using the equation (1)

$$\frac{1}{R_{tot}} = \frac{1}{R_{Ag}} + \frac{1}{R_{sc,dyn}} + \frac{1}{R_{subs}} \quad (1)$$

with the resistance of the silver stabilizer $R_{Ag}$, the substrate layer $R_{subs}$, and the variable dynamic resistance of the superconducting layer $R_{sc,dyn}$.

At direct current and no external magnetic field, the resistance in the superconductor $R_{sc,dyn}$ is zero and the current in the superconducting layer $I_{sc}$ corresponds to the transport current $I_t$. If an external magnetic field is applied and the dynamic resistance would be in the same order of magnitude as the other normal conducting resistances, the current is distributed across all layers. Therefore, the impact of increasing the dynamic resistance on the total resistance is highly

TABLE I
REBCO CONDUCTOR SPECIFICATIONS

| Manufacturer | SuperPower |
| --- | --- |
| Type | SF12100 |
| Minimal critical current at 77 K, s.f. | 338 A |
| Sample width | 12 mm |
| Thickness of Ag stabilizer layer each side | 1.0 µm |
| Thickness of superconductor layer | 1.0 µm |
| Thickness of substrate | 100 µm |
| Resistance per length at RT | 3.175 mΩ·cm$^{-1}$ |
| Resistance per length* at 77 K | 0.729 mΩ·cm$^{-1}$ |
| Critical temperature | 92 K |

*Resistance of tape at 77 K with superconducting layer at normal conducting conditions

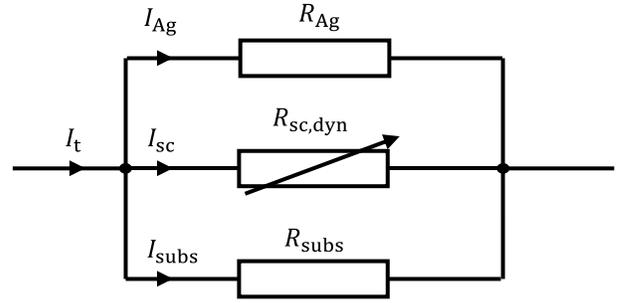

**Fig. 2.** Equivalent circuit of the multilayer superconductor

dependent on the normal conducting resistances in the superconducting tape.

For the investigated tape the resistance per length at 77 K of the silver stabilizer $R_{Ag}/l$ is 0.785 mΩcm$^{-1}$ and for the substrate layer $R_{subs}/l$ 10.25 mΩcm$^{-1}$ resulting in a parallel resistance $R_{nc}/l = R_{Ag}/l \| R_{subs}/l$ of 0.729 mΩcm$^{-1}$ for all normal conducting components. Despite the low thickness of the silver stabilizer, this has a high impact on the total resistance $R_{tot}$. Therefore, this will be further investigated in this work. The silver stabilizer is modified in two ways according to Figure 3. The silver is removed chemically with a solution of 50 % water, 25 % hydrogen peroxide, and 25 % ammonia. Afterward cleaned with distilled water and ethanol. The silver sections which should not be removed were masked by polyimide tape. Sample A is the reference sample without any modifications. In sample B the surrounding silver is removed in a 5 cm long section of the superconducting tape where the magnetic field will later be applied. This removes the parallel resistance of the silver completely. Sample C shows a tape with a remaining cap layer on the superconducting layer except for a 5 cm cutout in the middle. This eliminates the low-resistance connection from the superconducting layer to the substrate via the silver stabilizer.

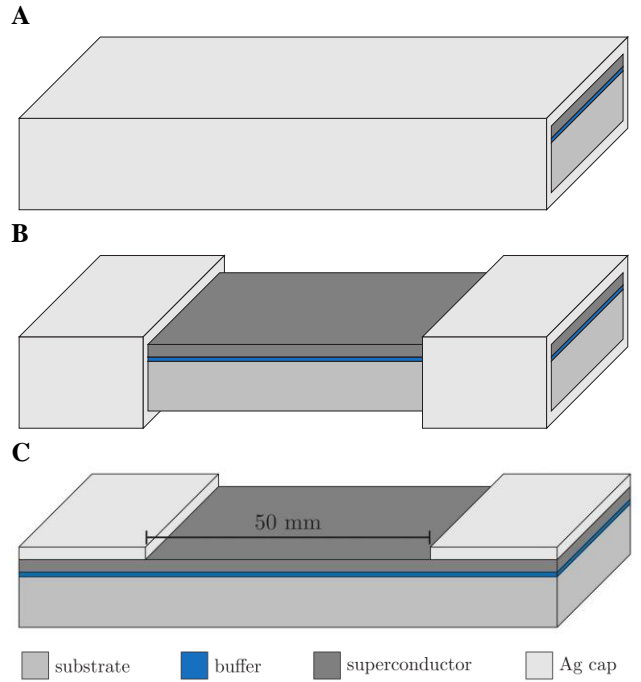

**Fig. 3.** Various etching patterns, whose effect on dynamic resistance were studied in more detail.

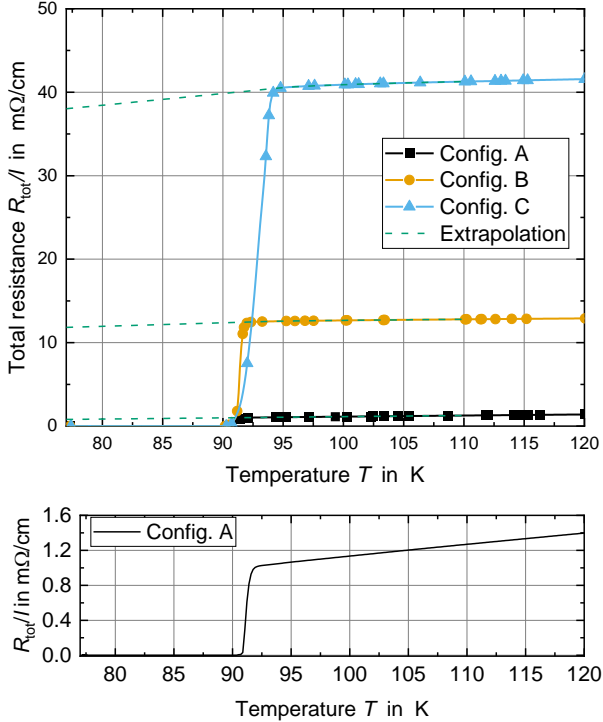

**Fig. 4.** Measured temperature-dependent resistance per length of the different tape configurations.

To see the impact on the resistance of such modifications, the temperature-dependent resistance $R(T)$ was measured for those samples from 77 K to 120 K and is shown in Figure 4. As expected, the silver layer geometry impacts the resistance significantly. A sharp transition between the superconducting and normal conducting zones at 92 K can be seen, although the transition zone is wider for configuration C. At 95 K the resistance per length is 1.06 mΩcm$^{-1}$ for configuration A, 12.6 mΩcm$^{-1}$ for configuration B, and 40.5 mΩcm$^{-1}$ for configuration C. At 120 K the resistance values are 1.4 mΩcm$^{-1}$, 12.9 mΩcm$^{-1}$ and 41.6 mΩcm$^{-1}$. These values are consistent with the parallel connection of all normal conducting resistances according to the thickness from Table 1 and temperature-dependent resistances in [33], [34], [35]. The normal conducting resistance was increased by a factor of 12 for configuration B and 40 for C.

Multiple critical current measurements before and after the removal of silver were made and no degradation of the critical current occurred. The electric field of 20µVcm$^{-1}$ was exceeded by increasing the transport current without any damage to the tape showing sufficient cooling by pool boiling. Additionally, tapes B and C were cycled thermally, and no degradation was observed. For environmental protection of the B and C tapes, a Kapton® foil with a thickness of 50 µm was laminated on the etched part of the tape, avoiding contact between the superconducting layer and air or liquid nitrogen.

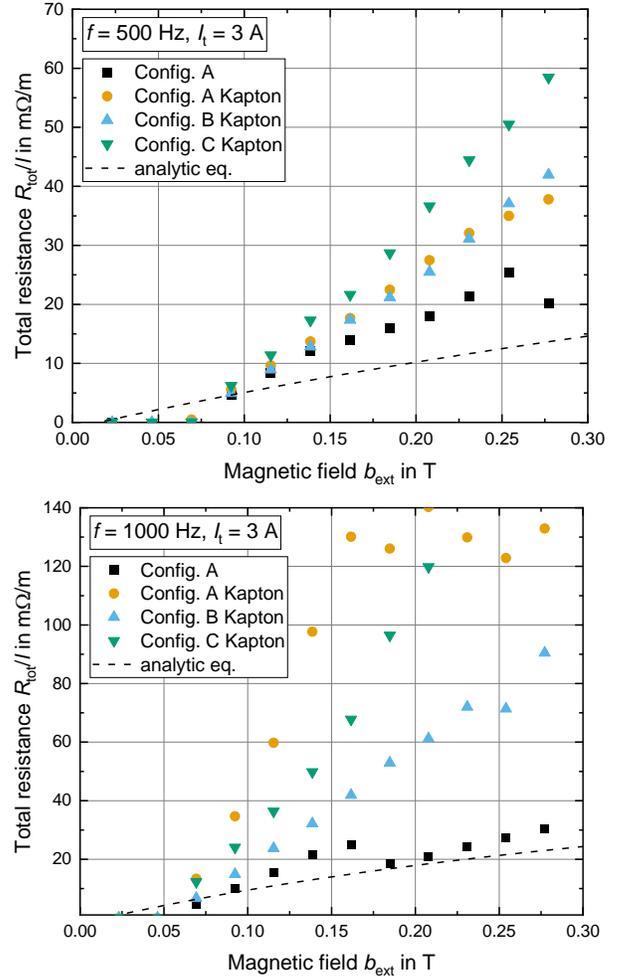

**Fig. 5.** Measured total resistance of various tape configurations according to Fig.3 for different external field amplitudes at 500 Hz and 1000 Hz.

## III. MEASUREMENTS RESULTS

The resistance for various configurations according to Figure 3 was measured and is shown in Figure 5. For comparison, all probes were laminated in Kapton®. Additionally, configuration A is displayed without a Kapton® protection layer. The figure shows the directly measured resistance, i.e. the total resistance, under varying the external magnetic field. Also, the analytical linear equation [36] with the parameter of the used superconductor is displayed. The data is presented for two different frequencies, 500 Hz (top) and 1000 Hz (bottom) at a fixed transport current of 3 A and the magnetic field varied from 25 mT to 277 mT. The plots show, that a threshold field amplitude $B_{th}$ exists, below which the dynamic resistance of the superconductor remains zero.

For 500 Hz, the resistances at lower magnetic fields up to 100 mT are very similar for all configurations, and they increase linearly with magnetic field. All Kapton® laminated configuration show a more significant rise compared to configuration A. With further increasing the magnetic field, the resistance values then diversify, with configuration A without Kapton® having the lowest values and configuration C with Kapton® having the highest values up to 58 mΩ·m$^{-1}$. At higher magnetic field, all resistances show a slight non-linear increase. Configuration A without protective Kapton® layer shows a decrease in resistance at about 0.25 T.

At 1000 Hz, the resistance increases linearly in the beginning and spreading further out with higher magnetic field. Configuration A without Kapton® also has a dip around 25 mΩ·m$^{-1}$ at the same point as with 500 Hz, and then increasing linearly afterward. This is further investigated in the next chapter. This configuration has the lowest resistance values and configuration A with Kapton the highest. Above 150 mT the total resistance reaches an upper ceiling of around 130 mΩ m$^{-1}$. Comparing this to the temperature dependent resistance shown in Fig. 3, the measured resistance is much higher than expected at 77 K. It can be assumed, that the temperature exceeds the critical temperature of the superconductor according to the temperature dependent resistance measurement the temperature of the superconductor at 1000 Hz and above 150 mT must be between 105 K and 120 K. This is further investigated with numerical calculations.

## IV. NUMERICAL COMPARISON

Additional to the measurement, multiple modelling approaches have been investigated to verify the measurement results. As already shown, the deviation between analytical equation and measurement is clearly visible. It has already been shown in [23], [24], [26], [27], [29] that the linear equation is accurate for low frequencies up to 100 Hz and low magnetic field amplitudes up to 100 mT. In this case the frequency is 5 to 10 times higher, and the influence of non-superconducting layers cannot be neglected [37] and basic analytical formulas are reaching their limits. Consequently, a numerical multilayer model [37] based on the H-formulation has been built and compared to the measurement, linear equation [36], [38] and nonlinear equation [16]. The superconductor parameters are listed in Table 1. The magnetic field dependence of the critical current is modelled with an elliptical equation:

$$I_c(B_\parallel, B_\perp) = \frac{I_{c0}}{\left(1 + \sqrt{(kB_\parallel)^2 + B_\perp^2}/B_c\right)^b} \quad (2)$$

where the parameters are $I_{c0} = 338$ A, $B_c = 42.65$ mT, $k = 0.29515$ and $b = 0.7$. No temperature dependence is included, therefore the temperature is constant at 77 K. The resistivities of the normal conducting layers are 2.5 nΩm for silver [33], [34] and 1.23 µΩm for substrate [35]. The air domain has been modeled by a resistivity of 2 Ωm. A perpendicular external magnetic field is applied varying from 25 mT to 350 mT while a transport current of 3 A is applied with boundary conditions. The transport current $I_t$ is the sum of all currents of each layer, $I_t = I_{sc} + I_{subs} + I_{Ag}$.

The simulation time consists of one full period where the second half-cycle is assumed as steady state and therefore used for the loss calculation. In general, the losses in a superconducting tape can be distinguished by the power source, current transport losses and magnetization losses. The total loss is the sum of these individual losses of each layer.

$$Q_{tot.layer} = Q_{mag,layer} + Q_{trans,layer} \quad (3)$$

The total losses per meter of each layer $Q_{tot.layer}$ is calculated by the integration of the joule heating over surface area of each layer.

$$Q_{tot.layer} = \iint EJ ds \quad (4)$$

where E is the electric field, J the current density and s the surface area of each layer. To distinguish both losses, the transport loss is calculated first with equation (4)

$$Q_{trans.layer} = I_{layer} \cdot \bar{E} \quad (5)$$

where $I_{layer}$ is the transport current of each layer and $\bar{E}$ the average voltage drop per length of the superconducting layer calculated by $\bar{E} = \iint E \, ds_{sc}/s_{sc}$ where $s_{sc}$ is the surface area of the superconducting layer. With both, the total loss $Q_{tot.layer}$ and the transport loss $Q_{trans.layer}$ the magnetization loss $Q_{mag.layer}$ can be calculated with

$$Q_{mag.layer} = Q_{tot.layer} - Q_{trans.layer} \quad (6)$$

The result of the modelling is pictured in Figures 6 to 9 for two different frequencies, 500 Hz and 1000 Hz, at a constant transport current of 3 A.

Figure 6 displays the total resistance per length $R_{tot}/l$ as a function of the external magnetic field $b_{ext}$. It compares the numerical results with the analytic non-linear equation [16] and the measurement results of configuration A. At both 500 Hz and 1000 Hz, the numerical data follows the corresponding analytical equation at lower magnetic fields. At higher magnetic fields both deviate from each other whereby the numerical simulation leads to higher resistances. The measurements however show higher values at lower magnetic amplitudes at higher fields the measurements follow closely the numeric simulation. This is the case above 277 mT for 500 Hz and 175 mT for 1000 Hz. This behavior is due to the transition of heat transfer mechanisms in liquid nitrogen and therefore influencing the heat flux from tape to fluid [39], [40]. This is further investigated in the next section.

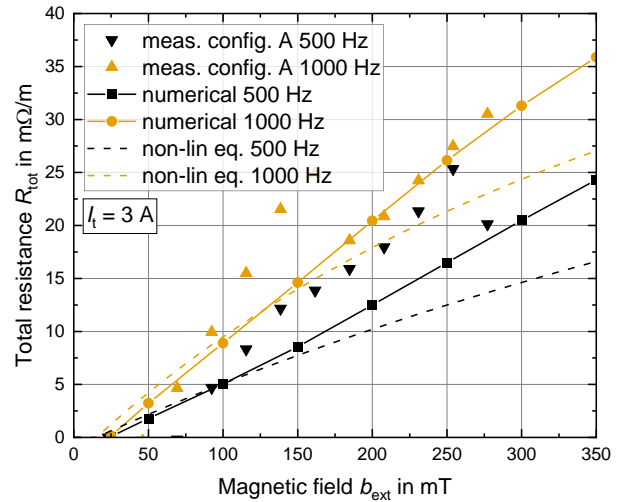

**Fig. 6.** Numerical modelling of total resistance per length compared to the analytical non-linear equation.

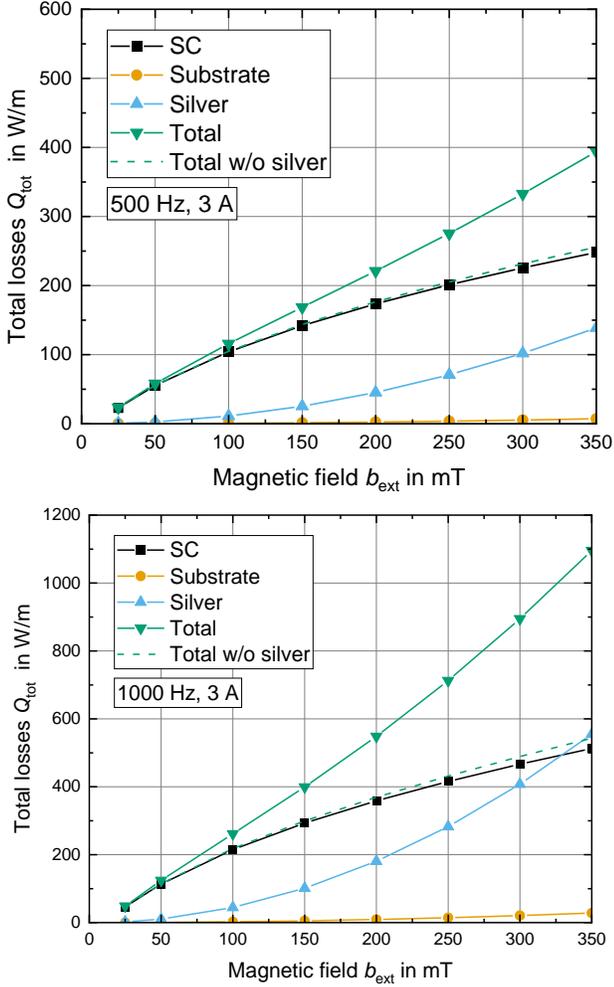

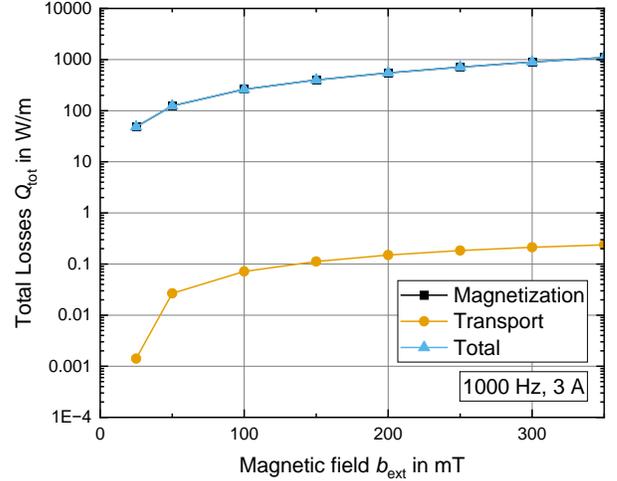

**Fig. 8.** Numerical modelling of losses divided in magnetization losses and current transport losses in a superconducting tape.

Figure 8 shows the total loss of the whole superconducting tape divided into transport losses and magnetization losses for a frequency of 1000 Hz and a transport current of 3 A. The transport losses show a steep increase at the beginning but level off with beyond 100 mT. These are significantly smaller than the magnetization losses. In this case, the magnetization losses dominate the overall losses generated in the superconducting tape.

The losses are dissipated via pool boiling in liquid nitrogen at ambient pressure. The heat flux density $\dot{q}$ in W/cm$^2$ can be calculated from the geometry of the superconductor and the generated losses and is shown for all modeled cases in Figure 9. The heat flux density is provided for two frequencies 500 Hz and 1000 Hz with and without the silver layers. For 500 Hz a maximum heat flux of 1.6 W/cm$^2$ is expected whereas at 1000 Hz the maximum heat flux density increases to 4.55 W/cm$^2$.

Applying these results to the boiling curve of liquid nitrogen from literature [39], [40], the excess temperature $\Delta T$ of the superconducting tape can be determined. At the setpoint of 250 mT and 1000 Hz with silver the heat flux density is approx. 3 W/cm$^2$. This means an excess temperature of 4-5 K for an untreated surface and 30–40 K for the Kapton® laminated tape. The latter case agrees well with the measured resistances and the temperature-dependent resistance curve in Figures 4 and 5. This shows the high impact of temperature on the total resistance.

**Fig. 7.** Numerical modelling of losses in different layers of a superconducting tape with and without silver layer for 500 Hz and 1000 Hz. Transport current is 3 A.

Figure 7 shows the total losses in different layers of the superconducting tape, with and without silver stabilizer, under varying external magnetic fields. In the case of 500 Hz, the losses of the whole tape increase steadily with the magnetic field. The contribution of the substrate layer is negligible, remaining flat across the entire range of external magnetic field. The losses in the silver layer grow quadratically with the magnetic field. The superconducting layer contributes significantly to the losses, but the gradient decreases as the field increases. The total loss without silver shows that removing silver has no significant impact on the losses in the other layers by following the slope of the superconducting layers with an offset added by the substrate.

In case of 1000 Hz, the losses are much higher compared to 500 Hz showing a non-linear dependency on the frequency. The silver layer's contribution increases sharply as the external magnetic field rises, exceeding the losses in the superconducting layer, making it the dominant factor in the total losses. The total loss in the superconducting layer scales linearly with the frequency. Again, the substrate layer is negligible. The total loss without silver is again significantly lower than with silver.

## V. CONCLUSION

The resistance of several HTS tape configurations were investigated over a wide range of frequencies and amplitudes up to 1000 Hz and 277 mT. The structure of the normal conducting layers of a superconductor was varied to achieve a higher resistance. As preliminary experiment, the temperature dependent resistance $R(T)$ has been measured from 77 K to 120 K showing the potential resistance increase for each tape configuration. With the highest increase of resistance with configuration C by a factor of 40 compared to the standard tape.

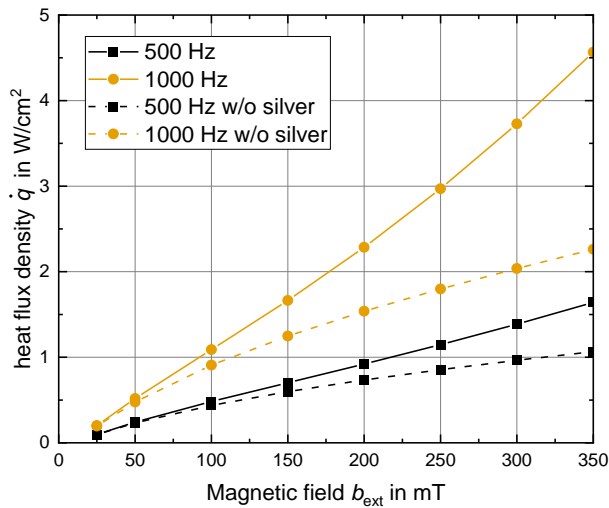

**Fig. 9.** Numerical modelling of heat flux density for cooling the superconducting tape.

The experimentally measured total resistances confirm this and show the increase in resistance with configurations B and C. However, for high frequencies the Kapton® laminated configuration A exceeded the other configurations, giving an increase by a factor of 4.3 compared to the standard unmodified superconducting tape. This phenomenon has been further investigated by a numerical model based on the H-formulation, since this behavior could not be fully explained by the normal conducting resistances of the superconducting tape. As results the losses in each layer of the superconductor were displayed. Simulations were performed on a tape with and without silver stabilizer from 25 mT to 350 mT external field. The simulation showed that the silver layer has a high impact on the total losses and therefore also on the joule heating of the whole tape. Combined with literature data of pool boiling in liquid nitrogen, an increase of the tape temperature by up to 40 K is possible. This corresponds well to the measured resistance and the temperature dependent resistance curve and show the high impact of the temperature on the built-up resistance.

With this result, superconducting switching devices will be investigated. A fully superconducting H-bridge inverter with 4 HTS switches will be built enabling dc /ac conversion at cryogenic conditions. The high resistances measured here are promising for obtaining fast commutation times and small leakage currents.

REFERENCES

[1] A. Kudymow, M. Noe, C. Schacherer, H. Kinder, and W. Prusseit, "Investigation of YBCO Coated Conductor for Application in Resistive Superconducting Fault Current Limiters," *IEEE Trans. Appl. Supercond.*, vol. 17, no. 2, pp. 3499–3502, Jun. 2007, doi: 10.1109/TASC.2007.899578.

[2] M. Noe *et al.*, "Conceptual Design of a 110 kV Resistive Superconducting Fault Current Limiter Using MCP-BSCCO 2212 Bulk Material," *IEEE Trans. Appl. Supercond.*, vol. 17, no. 2, pp. 1784–1787, Jun. 2007, doi: 10.1109/TASC.2007.898125.

[3] J. Zhu *et al.*, "Experimental investigation of current limiting characteristics for a novel hybrid superconducting fault current limiter (SFCL) with biased magnetic field," *J. Phys. Conf. Ser.*, vol. 1559, no. 1, p. 012104, Jun. 2020, doi: 10.1088/1742-6596/1559/1/012104.

[4] S. Hellmann, M. Abplanalp, S. Elschner, A. Kudymow, and M. Noe, "Current Limitation Experiments on a 1 MVA-Class Superconducting Current Limiting Transformer," *IEEE Trans. Appl. Supercond.*, vol. 29, no. 5, pp. 1–6, Aug. 2019, doi: 10.1109/TASC.2019.2906804.

[5] C. Schacherer *et al.*, "SmartCoil - Concept of a Full-Scale Demonstrator of a Shielded Core Type Superconducting Fault Current Limiter," *IEEE Trans. Appl. Supercond.*, vol. 27, no. 4, pp. 1–5, Jun. 2017, doi: 10.1109/TASC.2016.2642139.

[6] W. T. B. de Sousa, M. Noe, S. Huwer, and W. Reiser, "Design of a 110-kV 2.0-kA SmartCoil Superconducting Fault Current Limiter," *IEEE Trans. Appl. Supercond.*, vol. 33, no. 4, pp. 1–9, Jun. 2023, doi: 10.1109/TASC.2023.3246818.

[7] D. K. Park *et al.*, "Design and Test of a Thermal Triggered Persistent Current System using High Temperature Superconducting Tapes," *J. Phys. Conf. Ser.*, vol. 43, pp. 5–8, Jun. 2006, doi: 10.1088/1742-6596/43/1/002.

[8] S. B. Kim *et al.*, "Current Bypassing Properties by Thermal Switch for PCS Application on NMR/MRI HTS Magnets," *Phys. Procedia*, vol. 65, pp. 149–152, Jan. 2015, doi: 10.1016/j.phpro.2015.05.088.

[9] P. C. Michael, T. Qu, J. Voccio, J. Bascuñán, S. Hahn, and Y. Iwasa, "A REBCO Persistent-Current Switch (PCS): Test Results and Switch Heater Performance," *IEEE Trans. Appl. Supercond.*, vol. 27, no. 4, pp. 1–5, Jun. 2017, doi: 10.1109/TASC.2017.2652303.

[10] W. Li *et al.*, "Performance of a Persistent Current Switch for Large-Scale HTS Magnet," *IEEE Trans. Appl. Supercond.*, vol. 34, no. 8, pp. 1–4, Nov. 2024, doi: 10.1109/TASC.2024.3420319.

[11] T. Tosaka, T. Kuriyama, M. Yamaji, K. Kuwano, M. Igarashi, and M. Terai, "Development of a persistent current switch for HTS magnets," *IEEE Trans. Appl. Supercond.*, vol. 14, no. 2, pp. 1218–1221, Jun. 2004, doi: 10.1109/TASC.2004.830534.

[12] J. Ma *et al.*, "A Numerical Design of High-Resistance and Energy-Efficient HTS Switch Based on Dynamic Resistance," *IEEE Trans. Appl. Supercond.*, vol. 33, no. 5, pp. 1–5, Aug. 2023, doi: 10.1109/TASC.2023.3252493.

[13] J. Gawith, J. Geng, J. Ma, B. Shen, C. Li, and T. A. Coombs, "HTS Transformer–Rectifier Flux Pump Optimization," *IEEE Trans. Appl. Supercond.*, vol. 29, no. 5, pp. 1–5, Jan. 2019, doi: 10.1109/TASC.2019.2904444.

[14] B. Leuw, J. Geng, J. H. P. Rice, D. A. Moseley, and R. A. Badcock, "A half-wave superconducting transformer-rectifier flux pump using $J_c$ (B) switches," *Supercond. Sci. Technol.*, vol. 35, no. 3, p. 035009, Mar. 2022, doi: 10.1088/1361-6668/ac4f3d.

[15] Z. Jiang *et al.*, "The dynamic resistance of YBCO coated conductor wire: effect of DC current magnitude and applied field orientation," *Supercond. Sci. Technol.*, vol. 31, no. 3, p. 035002, Jan. 2018, doi: 10.1088/1361-6668/aaa49e.

[16] H. Zhang et al., "A full-range formulation for dynamic loss of high-temperature superconductor coated conductors," *Supercond. Sci. Technol.*, vol. 33, no. 5, p. 05LT01, May 2020, doi: 10.1088/1361-6668/ab7b0d.

[17] J. Ma, J. Geng, W. K. Chan, J. Schwartz, and T. Coombs, "A temperature-dependent multilayer model for direct current carrying HTS coated-conductors under perpendicular AC magnetic fields," *Supercond. Sci. Technol.*, vol. 33, no. 4, p. 045007, Apr. 2020, doi: 10.1088/1361-6668/ab6fe9.

[18] H. Zhang, C. Hao, Y. Xin, and M. Mueller, "Demarcation Currents and Corner Field for Dynamic Resistance of HTS-Coated Conductors," *IEEE Trans. Appl. Supercond.*, vol. 30, no. 8, pp. 1–5, Dec. 2020, doi: 10.1109/TASC.2020.3002209.

[19] H. Zhang, M. Yao, Z. Jiang, Y. Xin, and Q. Li, "Dependence of Dynamic Loss on Critical Current and $n$-Value of HTS Coated Conductors," *IEEE Trans. Appl. Supercond.*, vol. 29, no. 8, pp. 1–7, Dec. 2019, doi: 10.1109/TASC.2019.2948993.

[20] H. Zhang, P. Machura, K. Kails, H. Chen, and M. Mueller, "Dynamic loss and magnetization loss of HTS coated conductors, stacks, and coils for high-speed synchronous machines," *Supercond. Sci. Technol.*, vol. 33, no. 8, p. 084008, Aug. 2020, doi: 10.1088/1361-6668/ab9ace.

[21] Y. Sun, J. Geng, R. A. Badcock, and Z. Jiang, "Dynamic resistance and voltage response of a REBCO bifilar stack under perpendicular DC-biased AC magnetic fields," *Supercond. Sci. Technol.*, vol. 36, no. 9, p. 095014, Sep. 2023, doi: 10.1088/1361-6668/ace8c6.

[22] Z. Jiang et al., "Dynamic Resistance Measurement of a Four-Tape YBCO Stack in a Perpendicular Magnetic Field," *IEEE Trans. Appl. Supercond.*, vol. 28, no. 4, pp. 1–5, Jan. 2018, doi: 10.1109/TASC.2017.2787178.

[23] Z. Jiang, R. Toyomoto, N. Amemiya, C. W. Bumby, R. A. Badcock, and N. J. Long, "Dynamic Resistance Measurements in a GdBCO-Coated Conductor," *IEEE Trans. Appl. Supercond.*, vol. 27, no. 4, pp. 1–5, Jan. 2017, doi: 10.1109/TASC.2016.2644107.

[24] Z. Jiang, R. Toyomoto, N. Amemiya, X. Zhang, and C. W. Bumby, "Dynamic resistance of a high-$T_c$ coated conductor wire in a perpendicular magnetic field at 77 K," *Supercond. Sci. Technol.*, vol. 30, no. 3, p. 0301, Jan. 2017, doi: 10.1088/1361-6668/aa54e5.

[25] R. C. Duckworth, Y. F. Zhang, T. Ha, and M. J. Gouge, "Dynamic Resistance of YBCO-Coated Conductors in Applied AC Fields With DC Transport Currents and DC Background Fields," *IEEE Trans. Appl. Supercond.*, vol. 21, no. 3, pp. 3251–3256, Jan. 2011, doi: 10.1109/TASC.2010.2083621.

[26] Q. Li, M. Yao, Z. Jiang, C. W. Bumby, and N. Amemiya, "Numerical Modeling of Dynamic Loss in HTS-Coated Conductors Under Perpendicular Magnetic Fields," *IEEE Trans. Appl. Supercond.*, vol. 28, no. 2, pp. 1–6, Jan. 2018, doi: 10.1109/TASC.2017.2782712.

[27] J. M. Brooks, M. D. Ainslie, Z. Jiang, S. C. Wimbush, R. A. Badcock, and C. W. Bumby, "Numerical Modelling of Dynamic Resistance in a Parallel-Connected Stack of HTS Coated-Conductor Tapes," *IEEE Trans. Appl. Supercond.*, vol. 30, no. 4, pp. 1–8, Jan. 2020, doi: 10.1109/TASC.2020.2974860.

[28] M. D. Ainslie, C. W. Bumby, Z. Jiang, R. Toyomoto, and N. Amemiya, "Numerical modelling of dynamic resistance in high-temperature superconducting coated-conductor wires," *Supercond. Sci. Technol.*, vol. 31, no. 7, p. 074003, Jul. 2018, doi: 10.1088/1361-6668/aac1d3.

[29] J. M. Brooks, M. D. Ainslie, Z. Jiang, A. E. Pantoja, R. A. Badcock, and C. W. Bumby, "The transient voltage response of ReBCO coated conductors exhibiting dynamic resistance," *Supercond. Sci. Technol.*, vol. 33, no. 3, p. 035007, Jan. 2020, doi: 10.1088/1361-6668/ab6bfe.

[30] C. Li, Y. Xing, Y. Xin, B. Li, and F. Grilli, "Time-dependent development of dynamic resistance voltage of superconducting tape considering heat accumulation," *Superconductivity*, vol. 8, p. 100066, Dec. 2023, doi: 10.1016/j.supcon.2023.100066.

[31] J. Hu et al., "Numerical Study on Dynamic Resistance of an HTS Switch Made of Series-Connected YBCO Stacks," *IEEE Trans. Appl. Supercond.*, vol. 31, no. 5, pp. 1–6, Jan. 2021, doi: 10.1109/TASC.2021.3062258.

[32] J. Mun, C. Lee, C. Lee, K. Sim, and S. Kim, "An Experimental Study on the Dynamic Resistance of HTS Coil During Quasi-Persistent Current Operation Under External Harmonic Magnetic Field," *IEEE Trans. Appl. Supercond.*, vol. 34, no. 5, pp. 1–5, Aug. 2024, doi: 10.1109/TASC.2024.3374263.

[33] D. R. Smith and F. R. Fickett, "Low-Temperature Properties of Silver," *J. Res. Natl. Inst. Stand. Technol.*, vol. 100, no. 2, pp. 119–71, Jan. 1995, doi: 10.6028/jres.100.012.

[34] R. A. Matula, "Electrical resistivity of copper, gold, palladium, and silver," *J. Phys. Chem. Ref. Data*, vol. 8, no. 4, pp. 1147–1298, Jan. 1979, doi: 10.1063/1.555614.

[35] J. Lu, E. S. Choi, and H. D. Zhou, "Physical properties of Hastelloy ® C-276™ at cryogenic temperatures," *J. Appl. Phys.*, vol. 103, no. 6, p. 064908, Jan. 2008, doi: 10.1063/1.2899058.

[36] M. P. Oomen, J. Rieger, M. Leghissa, B. Haken, and H. H. J. Kate, "Dynamic resistance in a slab-like superconductor with $J_c(B)$ dependence," *Supercond. Sci. Technol.*, vol. 12, no. 6, pp. 382–387, Jan. 1999, doi: 10.1088/0953-2048/12/6/309.

[37] H. Zhang et al., "Modelling of electromagnetic loss in HTS coated conductors over a wide frequency band," *Supercond. Sci. Technol.*, vol. 33, no. 2, p. 025004, Jan. 2020, doi: 10.1088/1361-6668/ab6022.

[38] H. Zhang, B. Shen, X. Chen, and Z. Jiang, "Dynamic resistance and dynamic loss in a ReBCO superconductor," *Supercond. Sci. Technol.*, vol. 35, no. 11, p. 113001, Nov. 2022, doi: 10.1088/1361-6668/ac95d5.

[39] H. Merte Jr. and J. A. Clark, "Boiling Heat Transfer With Cryogenic Fluids at Standard, Fractional, and Near-Zero Gravity," *J. Heat Transf.*, vol. 86, no. 3, pp. 351–358, Aug. 1964, doi: 10.1115/1.3688689.

[40] S. Hellmann and M. Noe, "Influence of Different Surface Treatments on the Heat Flux From Solids to Liquid Nitrogen," *IEEE Trans. Appl. Supercond.*, vol. 24, no. 3, pp. 1–5, Jun. 2014, doi: 10.1109/TASC.2013.2283772.